\documentstyle[editedvolume,psfig]{crckapb}

\def\cl{\centerline}

\def\ea{\ et al. \,}
\def\eg{{\it e.g. \,}}
\def\be{\begin{equation}}
\def\ee{\end{equation}}
\def\rel{relativistic\,}

\begin{opening}
\title{NONTHERMAL PHENOMENA IN CLUSTERS OF GALAXIES}
\author{Yoel Rephaeli}
\institute{School of Physics \& Astronomy\\
Tel Aviv University, Tel Aviv, Israel}
\end{opening} 

\runningtitle{CLUSTER NT PHENOMENA} 
\begin{document}

\begin{abstract}

Recent observations of high energy ($> 20$ keV) X-ray emission in a few 
clusters extend and broaden our knowledge of physical phenomena in the 
intracluster space. This emission is likely to be nonthermal, probably 
resulting from Compton scattering of relativistic electrons by the cosmic 
microwave background radiation. Direct evidence for the presence of 
relativistic electrons in some $\sim 30$ clusters comes from measurements 
of extended radio emission in their central regions. I first review 
the results from RXTE and BeppoSAX measurements of a small sample of 
clusters, and then discuss their implications on the mean values of 
intracluster magnetic fields and \rel electron energy densities. 
Implications on the origin of the fields and electrons are briefly 
considered. 

\end{abstract}

\section{Introduction}

Extensive X-ray measurements of thermal bremsstrahlung emission from the hot, 
relatively dense intracluster (IC) gas, have significantly widened our 
knowledge of clusters of galaxies. Clusters are the largest bound systems in 
the universe, and as such constitute important cosmological probes. 
Their detailed dynamical and hydrodynamical properties are therefore 
of much interest. An improved understanding of the astrophysics of clusters, 
in particular, a more precise physical description of the IC environment, 
necessitates also adequate knowledge of the role of nonthermal phenomena.
The thermal state of IC gas may be appreciably affected by the presence of 
magnetic fields, and relativistic electrons and protons. Important processes 
involving the fields and particles are radio synchrotron emission, 
X-and-$\gamma$-ray emission from Compton scattering of electrons by the 
cosmic microwave background (CMB) radiation, nonthermal bremsstrahlung, decay 
of charged and neutral pions from proton-proton collisions, and gas heating 
by energetic protons. Detection of nonthermal X-ray emission from the 
electrons, when combined with radio measurements, yields direct information 
on the particles and fields, and establishes the basis for the study of 
nonthermal phenomena in clusters.

Clusters directly link phenomena on galactic and cosmological scales. As 
such, knowledge gained on magnetic fields and cosmic ray energy 
densities will form a tangible basis for the study of the origin of 
fields and (\rel) particles, and their distributions in the intergalactic 
space. Quantitative information on IC magnetic fields and \rel electrons 
is also very important for a realistic characterization of the processes 
governing their propagation in galactic halos and ejection to the IC space.

Nonthermal X-ray emission has recently been measured in a few clusters by 
the RXTE and BeppoSAX satellites. While we do not yet have detailed 
spectral and no spatial information on this emission, we are now able to 
determine more directly the basic properties of the emitting electrons and 
magnetic fields. Attempts to measure this emission will continue in the 
near future; spatial information could be obtained for the first time 
by observations with IBIS imager aboard the INTEGRAL satellite. In this 
short review, I describe the current status of the measurement of 
nonthermal X-ray emission by the RXTE and BeppoSAX satellites, and briefly 
discuss some of the implications on the properties of \rel electrons and 
magnetic fields in clusters. 

\section{Measurements}

\subsection{Radio Emission}

At present the main evidence for \rel electrons and magnetic fields in the 
IC space of clusters is provided by observations of extended radio emission 
which does not originate in the cluster galaxies. In a recent VLA survey of 
205 nearby clusters in the ACO catalog extended emission was measured in 32 
clusters (Giovannini \ea 1999, 2000). Only about a dozen of these were 
previously known to have regions of extended radio emission. In many of the 
clusters the emitting region is central, with a typical size of $\sim 1-3$ 
Mpc. The emission was measured in the frequency range $\sim 0.04-1.4$ GHz, 
with spectral indices and luminosities in the range $\sim$1--2, and 
$10^{40.5}$ -- $10^{42}$ erg/s ($H_0 = 50 \;km\;s^{-1} \;Mpc^{-1}$). 

Radio measurements yield a mean, volume-averaged field value of a few 
$\mu$G, under the assumption of global energy equipartition. 
The field can also be determined from measurements of Faraday rotation of 
the plane of polarization of radiation from cluster or background radio 
galaxies (\eg Kim \ea 1991). This has been accomplished statistically, by 
measuring the distribution of rotation measures (RM) of sources seen through 
a sample of clusters. In the recent study of Clarke \ea (2001) the width 
of the RM distribution in a sample of 16 nearby clusters was found to be 
about eight times larger than that of a control sample for radio sources 
whose lines of sights are outside the central regions of clusters. From the 
measured mean RM, field values of a few $\mu$G were deduced. 

It should be noted that the above methods to determine the field 
strength yield different spatial averages. Measurement of synchrotron 
emission yields a volume average of the field, whereas the measurement 
of Faraday rotation yields a line of sight average weighted by the 
electron density.

\subsection{Nonthermal X-ray Emission}

Compton scattering of the radio emitting (\rel) electrons by the CMB 
yields nonthermal X-ray and $\gamma$-ray emission. Measurement of
this radiation provides additional information that enables direct 
determination of the electron density and mean magnetic field, without 
the need to invoke equipartition (Rephaeli 1979). In the first 
systematic search for nonthermal X-ray emission in clusters, HEAO-1 
measurements of six clusters with regions of extended radio emission were 
analyzed (Rephaeli \ea 1987, 1988). The search continued with the CGRO
(Rephaeli \ea 1994), and ASCA (Henriksen 1998) satellites, but no significant
nonthermal emission was detected, resulting in lower limits on the mean, 
volume-averaged magnetic fields in the observed clusters, $B_{rx} 
\sim 0.1\; \mu$G. 

Significant progress in the search for this emission 
was recently made with the RXTE and BeppoSAX satellites. Evidence 
for the presence of a second component in the spectrum of the Coma cluster 
was obtained from RXTE ($\sim 90$ ks PCA, and $\sim 29$ ks HEXTE) 
measurements (Rephaeli \ea 1999). Although the detection of the second 
component was not significant at high energies, its presence was 
deduced at energies below 20 keV. Rephaeli \ea (1999) argued that this 
component is more likely to be nonthermal, rather than a second thermal 
component from a lower temperature gas. The two spectral components and 
the measurements are shown in Figure 1. The best-fit power-law photon 
index was found to be $2.3 \pm 0.45$ (90\% confidence), in good 
agreement with the radio index. The 2-10 keV flux in the power-law 
component is appreciable, $\sim 3\times 10^{-11}$ erg cm$^{-2}$ s$^{-1}$. 
The identification of the flux measured in the second component as Compton 
emission, when combined with the measured radio flux, yields 
$B_{rx}\sim 0.2\; \mu$G, and an electron energy density of $\sim 8 
\times 10^{-14} (R/1 Mpc)^{-3} $ erg cm$^{-3}$, scaling the radial extent 
(R) of the emitting region to 1 Mpc.
\begin{figure}
\cl{\psfig{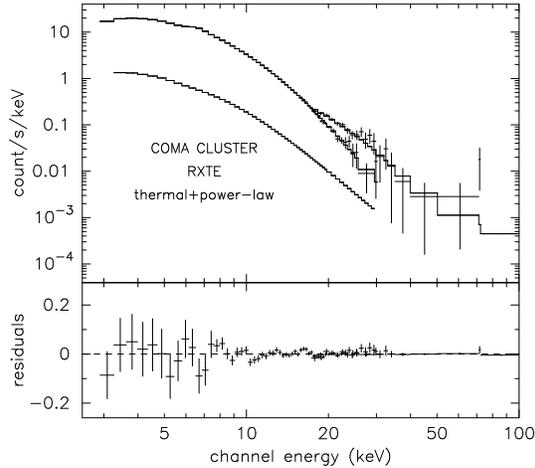}}
\caption{RXTE spectrum of the Coma cluster. Data and folded Raymond-Smith 
($kT \simeq 7.51$ keV), and power-law (index $=2.34$) models are shown in 
the upper frame; the latter component is also shown separately in the lower 
line. Residuals of the fit are shown in the lower frame.}        
\end{figure}

Observations of the Coma cluster with the PDS instrument aboard the 
BeppoSAX satellite have led to a direct measurement of a power-law 
component at high energies, $25-80$ keV (Fusco-Femiano 1999). A best-fit 
power-law photon index $2.6 \pm 0.4$ (90\% confidence) was deduced. The 
measured flux in a 20-80 keV band is $\sim 2\times 10^{-11}$ 
erg cm$^{-2}$ s$^{-1}$, about $\sim 8\%$ of the main, lower energy 2-10 
keV flux. A Compton origin yields $B_{rx}\sim 0.15\; \mu$G. These results 
are in good general agreement with the RXTE results (Rephaeli \ea 1999). 
In addition to Coma, power-law components were measured by BeppoSAX in 
two other clusters, A2199 (Kaastra \ea 2000), and A2256 (Fusco-Femiano 
2000). RXTE measurements of a second cluster, A2319, have also yielded
evidence for a second spectral component (Gruber \& Rephaeli 2001).
Note that A2199 is not known to have an extended region of IC radio emission.

While the measurements of second components in the spectra of these  
four clusters are significant, the identification of the emission as 
nonthermal is not certain. To better establish the nature of this emission, 
spatial information is needed in order to determine the location and size 
of the emitting region. The large fields of view of the RXTE ($\sim 1^{o}$) 
and BeppoSAX/PDS ($\sim 1.3^{o}$) instruments do not allow definite 
identification of the origin of the observed second spectral components. 
What seems to have been established, however, is that there is no 
temporal variability in the RXTE and BeppoSAX data, so an AGN 
origin of the second spectral components in these clusters is unlikely.

\section{Theory}

\subsection{Magnetic Fields}

The origin of IC magnetic fields is of interest in the study of the 
evolution of the IC environment and of fields on cosmological scales. 
IC fields could possibly be of cosmological origin -- intergalactic 
fields that had been enhanced during the formation and further evolution 
of the cluster, or else the fields might be mostly of galactic origin. The 
latter is a more likely possibility: fields were anchored to (`frozen in') 
the magnetized interstellar gas that was stripped from the member (normal 
and radio) galaxies (Rephaeli 1988). Dispersed galactic fields would have 
lower strengths and higher coherence scales in the IC space than is typical 
in galaxies. The field strength can be estimated under the assumption 
of flux-freezing, or magnetic energy conservation, if re-connection is 
ignored. Since typical galactic fields are $\sim$few $\mu$G, mean 
volume-averaged field values of a few $0.1 \mu$G are expected in the IC 
space. It is possible that Fields may have been amplified by the hydrodynamic 
turbulence generated by galactic motions (Jaffe 1980). However, this 
process was found to be relatively inefficient (Goldman \& Rephaeli 1991) 
in the context of a specific magneto-hydrodynamic model (Ruzmaikin \ea 
1989).

Estimates of magnetic fields in extended sources are usually based on 
measurements of the synchrotron emission by \rel electrons, and by 
Faraday rotation measurements. The estimates of $B_{rx}$ quoted 
in the previous section - based on radio synchrotron and Compton emission 
by what was {\it assumed} to be the same \rel electron population - are 
quite uncertain due to the need to make also other assumptions, such as 
the equality of the spatial factors in the expressions for the radio and 
X-ray fluxes (Rephaeli 1979). These are essentially volume integrations 
of the profiles of the electrons and fields, and since we have no 
information on the X-ray profile - and only rudimentary knowledge 
of the spatial distribution of the radio emission - this ratio was 
taken to be unity in the above mentioned analyses of RXTE and BeppoSAX 
data. The effect of this assumption is a systematically lower value of 
$B_{rx}$.
  
Faraday rotation measurements yield a weighted average of the field 
{\it and} gas density along the line of sight, $B_{fr}$. Estimates 
of $B_{fr}$ are also substantially uncertain, due largely to the 
fact that a statistically significant value of RM can only be obtained 
when the data from a sample of clusters are superposed (co-added). 
The significantly broader distribution of RM when plotted as function 
of cluster-centric distance clearly establishes the presence of 
IC magnetic fields. However, the deduced mean value of $B_{fr}$ is 
an average over all the clusters in the sample, in addition to 
being a weighted average of the product of a line of sight component 
of the field and the electron density. Both the field and density 
vary considerably across the cluster; in addition, the field is 
very likely tangled, with a wide range of coherence scales which 
can only be roughly estimated (probably in the range of $\sim 1-50$ kpc). 
All these make the determination of the field by Faraday rotation 
measurements considerably uncertain. 

The unsatisfactory observational status, and the intrinsic 
difference between $B_{rx}$ and $B_{fr}$, make it clear that 
these two measures of the field cannot be simply compared. 
Even ignoring the large observational and systematic uncertainties, 
the different spatial dependences of the fields, \rel electron 
density, and thermal electron density, already imply that 
$B_{rx}$ and $B_{fr}$ will in general be quite different. This 
was specifically shown by Goldshmidt \& Rephaeli (1993) in the context 
of reasonable assumptions for the field morphology, and the known 
range of IC gas density profiles. It was found that $B_{rx}$ is 
typically expected to be smaller than $B_{fr}$. Improved 
measurement of the spatial profile of the radio flux, and at 
least some knowledge of the spatial profile of the nonthermal X-ray 
emission, are needed before we can more meaningfully establish the 
relation between $B_{rx}$ and $B_{fr}$ in a given cluster.

\subsection{Energetic Particles}

The radio synchrotron emitting relativistic electrons in clusters 
lose energy also by Compton scattering. Where the field is $B < 3 
\;\mu$G, Compton losses dominate, and the characteristic loss time 
is $\tau_c \simeq 2.3/\gamma_3$ Gyr, where $\gamma_3$ is the Lorentz 
factor in units of $10^3$. Such electrons have Lorentz factors 
$\gamma_3 > 5$, if $B \sim 1 \mu$G, and therefore they lose their 
energy in less than 1 Gyr. If the mean field value is lower than 
$1 \mu$G, then the Compton loss time is even lower, because higher 
electron energies are needed to produce the observed radio emission. 
Clearly, if most of the radio-producing electrons had been injected 
from galaxies during a single, relatively short period, then the 
observed radio emission is a transient phenomenon, lasting only 
for a time $\sim \tau_c$, which is typically less than 1 Gyr. In this 
case the electron energy spectrum evolves on a relatively short 
timescale, shorter than typical evolutionary times of normal galaxies.
It follows that electrons from an early injection period would 
have lost their energy by now. If the emission is indeed relatively 
short-lived, then it may possibly be related to a few strong radio 
sources. 

On the other hand, if the observed radio emission has lasted more than 
a Compton loss time, then the \rel electron population has perhaps 
reached a (quasi) steady state. This could be attained through 
continual ejection of electrons from radio sources and other cluster 
galaxies, or by re-acceleration in the IC space. The lower the mean
value of the field, the shorter has the electron replenishment time to 
be. It is unclear whether any of these is a viable possibility, 
particularly so in the case of sub-$\mu$G fields. For electrons with 
energies $<< 1$ GeV, the main energy loss process is electronic 
(or Coulomb) excitations (Rephaeli 1979), and the loss time is maximal 
for $\gamma \sim 300$ (Sarazin 1999). In order for electrons with 
energies near this value to produce the observed IC radio emission, 
the mean magnetic field has to be at least $\sim$few $\mu$G. But in 
this case Compton scattering of these electrons by the CMB would only 
boost photon energies to the sub-keV range.

Various models have been proposed for acceleration of electrons in 
the IC space. All these invoke different scenarios of acceleration by  
shocks, resulting from galactic mergers, or shocks produced by fast 
moving cluster galaxies. In some of the accelerating electron models 
that have been proposed (Kaastra \ea 1998, Sarazin \& Kempner 2000), 
electrons produce (nearly) power-law X-ray emission by nonthermal 
bremsstrahlung. Presumably, all the radio, EUV, and X-ray measurements 
of Coma and A2199 can be explained as emission from a population of 
accelerating electrons (Sarazin \& Kempner 2000). However, the 
required electron energy density is much higher than in the \rel 
electron population that produces power-law X-ray emission by 
Compton scattering (Rephaeli 2001. Petrosian 2001). A more plausible 
acceleration model has been suggested by Bykov \ea (2000). 
In their model, galaxies with dark matter halos moving at supersonic 
(and super-Alfvenic) velocities can create collisionless bow shocks of 
moderate Mach number $M \geq$ 2. Kinetic modeling of nonthermal electron 
injection, acceleration and propagation in such systems seem to demonstrate 
that the halos are efficient electron accelerators, with the energy 
spectrum of the electrons shaped by the joint action of first and second 
order Fermi acceleration in a turbulent plasma with substantial Coulomb 
losses. Synchrotron, bremsstrahlung, and Compton losses of these electrons 
were found to produce spectra that are in quantitative agreement with 
current observations.

Although we do not yet have direct evidence for the presence of a 
substantial flux of energetic protons in the IC space, we do expect 
(based on the fact that protons are the main Galactic cosmic ray component) 
that they contribute very significantly - and even dominate - the cosmic 
ray energy density in clusters. Models for \rel IC protons (Rephaeli 1987) 
and their $\gamma$-ray emission by neutral pion decays (produced in p-p 
collisions), or by the radiation from secondary electrons resulting from 
charged pion decays, have been proposed (Dermer \& Rephaeli 1988, Blasi 
\& Colafrancesco 1999). A substantial low-energy proton component could 
also cause appreciable (Coulomb) heating of the gas in the cores of 
clusters (Rephaeli 1987, Rephaeli \& Silk 1995).

\section{Conclusion}

Recent RXTE and BeppoSAX measurements of power-law X-ray emission in four 
clusters improve our ability to characterize extragalactic magnetic fields 
and cosmic ray electrons. As expected, clusters of galaxies provide the 
first tangible basis for the exploration of nonthermal phenomena in 
intergalactic space. The radio and X-ray measurements strongly motivate 
further work on these phenomena. With the moderate spatial resolution 
(FWHM $\sim 12$ arcminute) capability of the IBIS imager on the INTEGRAL 
satellite (which is scheduled for launch in 2002) it will be possible 
to determine the morphology of nonthermal (at energies $>15$ keV) emission 
in nearby clusters. Such spatial information is crucial for a more definite 
identification of the observed second X-ray spectral components. 
Unequivocal measurement of cluster nothermal X-ray emission will 
greatly advance the study of nonthermal phenomena on cosmological 
scales.

\end{document}